\documentclass[12pt]{article}
\usepackage{epsfig}
\usepackage{axodraw}

\def\simlt{\stackrel{<}{{}_\sim}}

\newcommand{\bea}{\begin{eqnarray}}
\newcommand{\eea}{\end{eqnarray}}
\newcommand{\bd}{\begin{displaymath}}
\newcommand{\ed}{\end{displaymath}}
\newcommand{\be}{\begin{equation}}
\newcommand{\ee}{\end{equation}}

\renewcommand{\baselinestretch}{1.2}
\begin{document}

\thispagestyle{empty}

{\normalsize\sf
\rightline {hep-ph/0110237}
\rightline{IFT-01/28}
\vskip 3mm
\rm\rightline{October 2001}
}

\vskip 5mm

\begin{center}
  
{\LARGE\bf Low energy threshold corrections to neutrino masses 
and mixing angles}

\vskip 10mm

{\large\bf Piotr H. Chankowski and Pawe\l ~Wasowicz}\\[5mm]

Institute of Theoretical Physics, Warsaw University\\
Ho\.za 69, 00-681 Warsaw, Poland

\end{center}

\vskip 5mm

\renewcommand{\baselinestretch}{1.1} 
\begin{abstract}
\vskip 3mm 
We compute the low energy threshold corrections to neutrino masses and 
mixing in the Standard Model (SM) and its minimal supersymmetric version, 
using the effective theory technique. We demonstrate that they stabilize 
the renormalization group (RG) running with respect to the choice of the 
scale to which the RG equation is integrated. (This confirms the correctness
of the recent re-derivation of the RGE for the SM in hep-ph/0108005.) The 
explicit formulae for the low energy threshold corrections corrections can 
be applied to specific models of neutrino masses and mixing. 
\end{abstract}
\renewcommand{\baselinestretch}{1.2}

\newpage \setcounter{page}{1} \setcounter{footnote}{0}
%%%%%%%%%%%%%%%%%%%%%%%%%%%%%%%%%%%%%%%%%%%%%%%%%%%%%%%%%%%%%%%%%%%%%%%%%%%
% The main part of the paper                                              %
%%%%%%%%%%%%%%%%%%%%%%%%%%%%%%%%%%%%%%%%%%%%%%%%%%%%%%%%%%%%%%%%%%%%%%%%%%%
\section{Introduction}
\setcounter{equation}{0} 

There is at present a strong experimental evidence for neutrino oscillations.
Their most natural explanation is the existence of neutrino masses. Neutrino 
masses can be incorporated in the Standard Model (SM) or its supersymmetric 
extension (MSSM) by adding to the Lagrangian the non-renormalizable 
dimension-5 operator \cite{WE}
\begin{eqnarray}
\Delta {\cal L}_{\rm SM}=-{1\over4M}\mbox{\boldmath$C$}^{AB}
\left(\epsilon_{ki}H_kl_i^A\right)\left(\epsilon_{lj}H_ll_j^B\right)
+{\rm H.c.}.\label{eqn:dL_SM}
\end{eqnarray}
in which $A,B=1,2,3$ label generations, $l_j^A=(\nu^A, e^A)$ are 
the Weyl spinors transforming as doublets of $SU_L(2)$,  $H_i$ is the Higgs 
doublet with hypercharge $+1/2$ and $\epsilon_{21}=-\epsilon_{12}=1$, 
$\epsilon_{ii}=0$. After the electroweak symmetry breaking (\ref{eqn:dL_SM}) 
gives the neutrino mass matrix in the form
\begin{eqnarray}
\left(m_\nu^{\rm tree}\right)^{AB}={1\over4M}\mbox{\boldmath$C$}^{AB}v^2
\label{eqn:treemass}
\end{eqnarray}
(where $v$ is the vacuum expectation value of the neutral component of the
Higgs field $H_i$), which is diagonalized by the unitary rotation
$\nu_A\rightarrow U^{Aa}\nu_a$. The elements of the matrix $U$ determine the
neutrino oscillation probabilities and are, therefore, probed in the neutrino
experiments.

The operator (\ref{eqn:dL_SM}) appears in the low energy effective theory
as a result of integrating out fields of an underlying theory describing 
physics at some high energy scale. Thus, it is supposed to be generated at 
some high scale $M_F$ (much higher than the electroweak scale $M_Z$). 
Therefore obtaining reliable predictions for neutrino masses and mixing 
angles requires solving the renormalization group equation (RGE) for the 
Wilson coefficient $\mbox{\boldmath$C$}^{AB}$ \cite{CHPL,ANDRKELIRA} 
\begin{eqnarray}
{d\over dt}\mbox{\boldmath$C$}^{AB}=K\mbox{\boldmath$C$} ^{AB}
+\kappa\left[y^2_{e_A}\mbox{\boldmath$C$}^{AB} + \mbox{\boldmath$C$}^{AB}
y^2_{e_B}\right]\label{eqn:rge}
\end{eqnarray}
where $t=(1/16\pi^2)\ln(Q/M_Z)$ and
$y^2_{e_A}$ are the Yukawa couplings of the charged leptons.\footnote{Without 
loss of generality throughout the paper we work in the basis in which the 
matrix of the charged lepton Yukawa couplings is diagonal; the Wilson 
coefficient $\mbox{\boldmath$C$}^{AB}$ is therefore assumed to be given 
in that basis too. The normalization of $\lambda$ is fixed by the Higgs self 
interaction: ${\cal L}_{\rm self}=-{\lambda\over2}(H^\dagger H)^2$.} 
In the SM $\kappa=-3/2$ and $K=-3g^2_2+2\sum_{\rm fermions}N^{(f)}_cy^2_f$
where $N^{(f)}_c=3$ for quarks and $1$ for leptons; in the MSSM $\kappa=+1$ 
and $K=-6g^2_2 - 2g^2_Y + 6\sum_Ay^2_{u_A}$.

The solution to eq.~(\ref{eqn:rge})  \cite{ELLO}
\begin{eqnarray}
\mbox{\boldmath$C$}(Q)= I_K{\cal J}\mbox{\boldmath$C$}(M_F){\cal J}
\label{eqn:ellosol}
\end{eqnarray}
where ${\cal J}={\rm diag}(I_e,I_\mu,I_\tau)$ and 
\begin{eqnarray}
I_K=\exp\left(-\int_0^{t_Q}K(t^\prime)dt^\prime\right)
\phantom{aaaaa},\nonumber\\
I_{e_A}=\exp\left(-\kappa\int_0^{t_Q}y^2_{e_A}(t^\prime)dt^\prime\right)
\approx1-I^{\rm rg}_A,
\label{eqn:lolaIfact}
\end{eqnarray}
with $t_Q=(1/16\pi^2)\ln(M_F/Q)$,
gives $\mbox{\boldmath$C$}(Q)$ at the electroweak scale
$Q\approx M_Z$ in terms of $\mbox{\boldmath$C$}(M_F)$. The RGE (\ref{eqn:rge})
was analyzed in many papers \cite{ALL} to see how much the initial pattern
of neutrino masses and mixing angles generated at the scale $M_F$ is modified
by quantum corrections involving large logarithms $\ln(M_F/M_Z)\gg1$. In 
particular, it has been found \cite{CHKRPO} that eq.~(\ref{eqn:rge}) exhibits
a nontrivial fixed point structure. In many interesting cases (e.g. for 
degenerate or partially degenerate neutrino mass spectrum) that structure 
leads to the 
pattern of mixing angles that is not compatible with the present experimental 
indications (i.e. bimaximal mixing and small $U_{13}$ matrix element).
It has been however pointed out \cite{CHUPO,CHIOPOVA,CHU} that in the MSSM
the so-called low energy threshold corrections which were 
neglected in previous analyses \cite{ALL,CHKRPO} can in some cases be more 
important than the RG evolution and can change qualitatively the pattern 
obtained by solving eq.~(\ref{eqn:rge}).

In this paper we compute these low energy threshold corrections both in the 
SM and in the MSSM. We first show that they stabilize the results of the RG 
running with respect to the choice of the low energy scale $Q$ (to clarify  
the points raised in the recently  published paper \cite{SI} we demonstrate 
this explicitly in a pedagogical way) and asses their magnitude and dependence
on the parameters of the MSSM. 

\section{Standard Model}
\setcounter{equation}{0} 

In this Section we calculate one-loop corrections to the neutrino mass
matrix in the SM. Our starting point is the SM Lagrangian (see e.g. ref.
\cite{POKBOOK}) supplemented with the non-renormalizable term 
(\ref{eqn:dL_SM}). All parameters of this Lagrangian are understood to be 
running parameters renormalized at the scale $Q\sim M_Z$. Also
the Wilson coefficient $\mbox{\boldmath$C$}^{AB}$ of the dimension 5
$\Delta L=2$ operator (\ref{eqn:dL_SM}) is a renormalized parameter of
the effective theory Lagrangian. Integrating its RGE (\ref{eqn:rge}) from 
the high scale $M_F$ down to some scale $Q\approx M_Z$ resums
potentially large corrections involving $\ln(M_F/Q)$ to all orders of the 
perturbation expansion. However, since the low energy scale $Q$ is not 
{\sl a priori} determined by any physical 
requirement (apart from the condition $Q\sim M_Z$), the neutrino masses 
and mixing angles computed in the tree-level approximation from the Wilson
coefficient $\mbox{\boldmath$C$}^{AB}(Q)$ do depend (albeit weakly) on the
actual choice of $Q$. This dependence can be removed by computing masses and 
mixing angles in the one-loop approximation in the $\overline{\rm MS}$ 
scheme with the same renormalization scale $Q$. 

Since the neutrino masses are orders of magnitude smaller than the electroweak 
scale, the calculation of the low energy threshold corrections is technically 
most easily achieved in the effective theory approach. At the scale 
$Q\approx M_Z$ all 
gauge and Higgs bosons are integrated out and the effective theory valid below 
the electroweak scale is constructed. In this low energy theory the one-loop 
neutrino mass matrix $\left(m_\nu^{\rm 1-loop}\right)^{AB}$ is given by the 
tree-level term (\ref{eqn:treemass}) of the SM plus the one-loop (threshold) 
correction $\Delta m_\nu$.  The latter, apart from having leading $\ln Q$ 
dependence that exactly matches the $\ln Q$ dependence of the tree level mass 
$(m_\nu^{\rm tree})^{AB}(Q)$ (\ref{eqn:treemass}), can also contain nontrivial 
$Q$-independent pieces.

Writing the SM Higgs doublet as
\begin{eqnarray}
H={1\over\sqrt2}\left(\matrix{\sqrt2G^+\cr v+\phi^0+iG^0}\right)
\end{eqnarray}
we get from (\ref{eqn:dL_SM}) the neutrino mass term and various
interactions (we write down only those which will be relevant for us):
\begin{eqnarray}
&&\Delta {\cal L}_{\rm SM}=-{1\over2}(m_\nu^{\rm tree})^{AB}\nu_A\nu_B
\label{eqn:dL_SMint}\\
&&\phantom{aaaaaaa}+{v\mbox{\boldmath$C$}^{AB}\over2\sqrt2M}G^+e_B\nu_A
-{v\mbox{\boldmath$C$}^{AB}\over4M}\phi^0\nu_B\nu_A
-{\mbox{\boldmath$C$}^{AB}\over8M}(\phi^0\phi^0-G^0G^0)\nu_B\nu_A
+{\rm H.c.}\nonumber
\end{eqnarray}
In principle to get the Feynman rules for neutrino mass eigenstates $\nu_a$ 
one has also to rotate the neutrino fields $\nu_A\rightarrow U^{Aa}\nu_a$ 
(where $U^{Aa}$ is the matrix diagonalizing $(m_\nu^{\rm tree})^{AB}$). 
This step is however 
unnecessary for our purpose since we will use everywhere massless neutrino
propagators on internal lines.\footnote{Taking non-zero neutrino masses into
account in propagators would amount to including $1/M^2$ effects (where $M$
is the mass scale of the heavy neutrino states). Since we do not consider 
operators 
of dimension higher than five resulting from the seesaw mechanism we cannot
compute $1/M^2$ effects consistently.} We can therefore compute directly
the corrections to the tree level mass matrix $\mbox{\boldmath$C$}^{AB}$ in
the basis in which it is not necessarily diagonal.

\begin{figure}[htbp]
\begin{center}
%\begin{tabular}{lp{280\unitlength}}
\begin{picture}(300,100)(0,0)
\Vertex(70,50){13}
\ArrowLine(60,50)(20,50)
\ArrowLine(120,50)(80,50)
\Text(30,60)[]{$\nu_A$}
\Text(110,60)[]{$\nu_B$}
\Text(70,10)[]{$-i\Sigma_V^{AB}(p^2)\bar\sigma^\mu p_\mu$}
\Vertex(230,50){13}
\ArrowLine(180,50)(220,50)
\ArrowLine(280,50)(240,50)
\Text(190,60)[]{$\nu_A$}
\Text(270,60)[]{$\nu_B$}
\Text(230,10)[]{$-i\Sigma_m^{AB}(p^2)$}
\end{picture}
\end{center}
\caption{One-particle irreducible threshold corrections.}
\label{fig:1picorr}
%\end{tabular}
\end{figure}
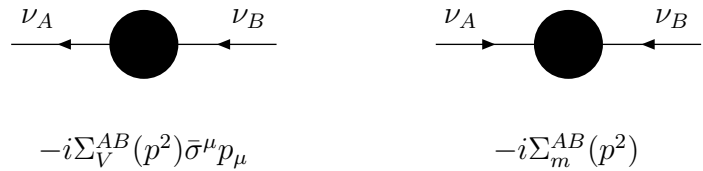

The strategy is now to integrate out heavy fields: $W^\pm$, $Z^0$, $\phi^0$
(as well as Goldstone bosons $G^\pm$ and $G^0$) and to construct the effective 
Lagrangian valid below the scale $Q\sim M_Z$. Up to terms of order 
${\cal O}(p^2/M^2_Z)$ where $p\sim m_\nu\ll M_Z$ is the external four momentum,
one-loop effects (shown schematically in fig.~\ref{fig:1picorr}) of the heavy 
fields present in the full SM have to be simulated by the corrections 
$\delta z^{AB}$ and $\delta m_\nu^{AB}$ in the effective theory Lagrangian
\begin{eqnarray}
{\cal L}_{\rm eff}=\bar\nu_A\left(\delta^{AB}+\delta z^{AB}\right)
i\bar\sigma^\mu\partial_\mu\nu_B - {1\over2}\left[\left(m_\nu^{\rm tree}
+\delta m_\nu\right)^{AB}\nu_A\nu_B+{\rm H.c.}\right] +\dots\label{eqn:leff}
\end{eqnarray}

Redefining the neutrino fields to get their kinetic term canonical 
and using $\delta z^{AB}=-\Sigma_V^{AB}(0)$,  
$\delta m_\nu^{AB}=\Sigma_m^{AB}(0)$
(where $\Sigma_V^{AB}(p^2)$ and $\Sigma_m^{AB}(p^2)$ are defined in 
fig.~\ref{fig:1picorr}) one gets
\begin{eqnarray}
&&\left(\Delta m_\nu\right)^{AB} = 
I^{\rm th}_{A^\prime A}\left(m_\nu^{\rm tree}\right)^{A^\prime B}
+\left(m_\nu^{\rm tree}\right)^{AB^\prime}I^{\rm th}_{B^\prime B}
\phantom{aaa}\label{eqn:Icordef}\\
&&\phantom{aaaaa}
I^{\rm th}_{AB}\equiv{1\over2}\Sigma_V^{AB}(0)+{1\over2}\Delta^{AB}\nonumber
\end{eqnarray}
where we have split 
\begin{eqnarray}
\Sigma_m^{AB}(0) = 
{1\over2}\Delta^{A^\prime A}\left(m_\nu^{\rm tree}\right)^{A^\prime B}
+{1\over2}\left(m_\nu^{\rm tree}\right)^{AB^\prime}\Delta^{B^\prime B}.
\end{eqnarray}

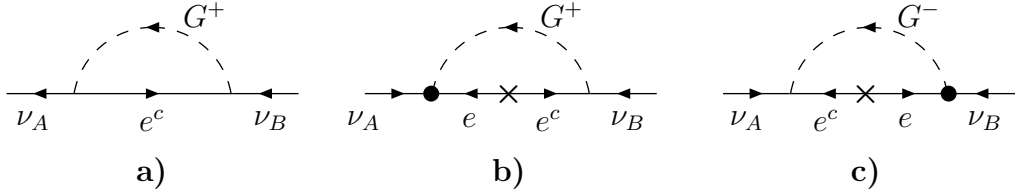
\begin{figure}[htbp]
\begin{center}
%\begin{tabular}{lp{280\unitlength}}
\begin{picture}(390,100)(0,0)
\ArrowLine(30,40)(5,40)
\ArrowLine(30,40)(90,40)
\ArrowLine(115,40)(90,40)
\DashArrowArc(60,35)(30,10,170){5}
\Text(15,30)[]{$\nu_A$}
\Text(105,30)[]{$\nu_B$}
\Text(60,30)[]{$e^c$}
\Text(80,70)[]{$G^+$}
\Text(60,10)[]{\bf a)}
\ArrowLine(140,40)(165,40)
\ArrowLine(195,40)(165,40)
\ArrowLine(195,40)(225,40)
\ArrowLine(250,40)(225,40)
\DashArrowArc(195,35)(30,10,170){5}
\Vertex(165,40){3}
\Text(195,40)[]{$\mbox{\boldmath$\times$}$}
\Text(140,30)[]{$\nu_A$}
\Text(240,30)[]{$\nu_B$}
\Text(180,30)[]{$e$}
\Text(210,30)[]{$e^c$}
\Text(215,70)[]{$G^+$}
\Text(195,10)[]{\bf b)}
\ArrowLine(275,40)(300,40)
\ArrowLine(330,40)(300,40)
\ArrowLine(330,40)(360,40)
\ArrowLine(385,40)(360,40)
\DashArrowArc(330,35)(30,10,170){5}
\Text(330,40)[]{$\mbox{\boldmath$\times$}$}
\Vertex(360,40){3}
\Text(285,30)[]{$\nu_A$}
\Text(375,30)[]{$\nu_B$}
\Text(315,30)[]{$e^c$}
\Text(345,30)[]{$e$}
\Text(350,70)[]{$G^-$}
\Text(330,10)[]{\bf c)}
\end{picture}
\end{center}
\caption{\protect Contributions of the charged Goldstone boson. Heavy dots 
indicate vertices arising from the operator (\ref{eqn:dL_SM}). Crosses
indicate fermion propagators with a helicity flip (i.e. with the fermion mass
in the numerator).}
\label{fig:cGcontr}
%\end{tabular}
\end{figure}

Goldstone bosons $G^\pm$ contribution to $\Sigma_V^{AB}$ and $\Sigma_m^{AB}$
are shown in figs.~\ref{fig:cGcontr}a and b, c, respectively, where the dots 
denote interaction vertex originating from $\Delta{\cal L}_{\rm SM}$ given in 
eq.~(\ref{eqn:dL_SMint}). In a general $R_\xi$-gauge and working in the
$\overline{\rm MS}$ scheme one finds
\begin{eqnarray}
&&\Sigma_V^{AB}(0) = {1\over2}\delta^{AB}y^2_{e_A}
\left[(m^2_{e_A}-\xi_WM^2_W)B^\prime_0(e_A,G^\pm)+B_0(e_A,G^\pm)\right]
\nonumber\\
&&\Sigma_m^{AB}(0) = -{v^2\over4M}\left[y^2_{e_A}\mbox{\boldmath$C$}^{AB}
B_0(e_A,G^\pm)+\mbox{\boldmath$C$}^{AB}y^2_{e_B}B_0(e_B,G^\pm)\right].
\phantom{aa}
\end{eqnarray}
We have used the abbreviated notation for the standard two-point function 
\begin{eqnarray}
B_0(1,2)\equiv B_0(0,m_1,m_2) ={1\over(4\pi)^2}\left[-1
+{m^2_1\over m^2_1-m^2_2}\ln{m^2_1\over Q^2}
+{m^2_2\over m^2_2-m^2_1}\ln{m^2_2\over Q^2}\right]\phantom{aaaa} \nonumber\\
B^\prime_0(1,2)\equiv 
\left.{d\over dp^2}B_0(p^2,m_1,m_2)\right|_{p^2=0} 
={1\over(4\pi)^2}\left[-{1\over2}{m^2_1+m^2_2\over(m^2_1-m^2_2)^2}+
{m^2_1m^2_2\over(m^2_1-m^2_2)^3}\ln{m^2_1\over m^2_2}\right]\nonumber
\end{eqnarray}
where $m_1$ and $m_2$ are the masses of particles 1 and 2.

The $W^\pm$ boson exchange contributes only to $\Sigma_V$:
\begin{eqnarray}
&&\Sigma_V^{AB}={g_2^2\over2}\delta^{AB}
\left[(m^2_{e_A}-M^2_W)B^\prime_0(e_A,W^\pm)+B_0(e_A,W^\pm)+1\right]
\nonumber\\
&&\phantom{aaaa}+{g_2^2\over4}\delta^{AB}
\left\{{m^2_{e_A}\over M^2_W}
\left[(m^2_{e_A}-M^2_W)B^\prime_0(e_A,W^\pm)-B_0(e_A,W^\pm)\right]
-2B_0(e_A,W^\pm)\right.\phantom{aaa}\\
&&\phantom{aaaaaaaaaa}\left.
-{m^2_{e_A}\over M^2_W}
\left[(m^2_{e_A}-\xi_WM^2_W)B^\prime_0(e_A,G^\pm)-B_0(e_A,G^\pm)\right]
+2\xi_WB_0(e_A,G^\pm)\right\}\nonumber
\end{eqnarray}

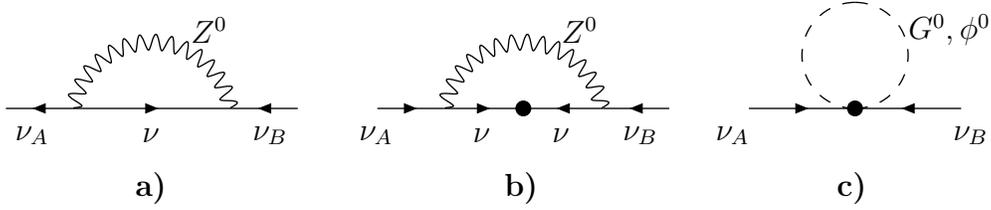
\begin{figure}[htbp]
\begin{center}
%\begin{tabular}{lp{280\unitlength}}
\begin{picture}(390,100)(0,0)
\ArrowLine(35,40)(10,40)
\ArrowLine(35,40)(95,40)
\ArrowLine(120,40)(95,40)
\PhotonArc(65,35)(30,10,170){3}{15}
\Text(20,30)[]{$\nu_A$}
\Text(110,30)[]{$\nu_B$}
\Text(65,30)[]{$\nu$}
\Text(87,70)[]{$Z^0$}
\Text(65,10)[]{\bf a)}
\ArrowLine(150,40)(175,40)
\ArrowLine(175,40)(205,40)
\ArrowLine(235,40)(205,40)
\ArrowLine(260,40)(235,40)
\PhotonArc(205,35)(30,10,170){3}{15}
\Vertex(205,40){3}
\Text(150,30)[]{$\nu_A$}
\Text(250,30)[]{$\nu_B$}
\Text(190,30)[]{$\nu$}
\Text(220,30)[]{$\nu$}
\Text(227,70)[]{$Z^0$}
\Text(205,10)[]{\bf b)}
\ArrowLine(290,40)(330,40)
\ArrowLine(370,40)(330,40)
\DashArrowArc(330,60)(20,90,450){5}
\Vertex(330,40){3}
\Text(285,30)[]{$\nu_A$}
\Text(375,30)[]{$\nu_B$}
\Text(367,70)[]{$G^0,\phi^0$}
\Text(330,10)[]{\bf c)}
\end{picture}
\end{center}
\caption{Contributions of $Z^0$ and neutral scalars.}
\label{fig:ZGcontr}
%\end{tabular}
\end{figure}

Together, $G^\pm$ and $W^\pm$ contributions give
\begin{eqnarray}
&&I_{AB}^{W^\pm G^\pm}=\delta^{AB}{g_2^2\over2}{m^2_{e_B}\over M_W^2}{1\over4}
\left[(m^2_{e_B}-M^2_W)B^\prime_0(e_B,W^\pm)-3B_0(e_B,W^\pm)\right]\nonumber\\
&&\phantom{aaaaa}
+\delta^{AB}{g_2^2\over4}
\left[(m^2_{e_B}-M^2_W)B^\prime_0(e_B,W^\pm)+B_0(e_B,W^\pm)+1\right]
\label{eqn:GWcontr}\\
&&\phantom{aaaaa}
-\delta^{AB}{g_2^2\over4}
\left[\left(1-{m^2_{e_B}\over M^2_W}\right)B_0(e_B,W^\pm)
-\left(\xi_W-{m^2_{e_B}\over M^2_W}\right)B_0(e_B,G^\pm)\right]
\phantom{aa}\nonumber
\end{eqnarray}

In the limit of massless neutrinos on internal lines, the contribution 
of $Z^0$ exchange to $\Sigma_V$ arising from the diagram shown in 
Fig.~\ref{fig:ZGcontr}a can by obtained from (\ref{eqn:GWcontr}) by setting 
there $m_{e_B}=0$ and replacing $W^\pm(G^\pm)\rightarrow Z^0(G^0)$, 
$g^2_2\rightarrow(g^2_2+g^2_Y)/2$. This gives 
\begin{eqnarray}
I^{Z^0(1)}_{AB}=\delta^{AB}{g^2_2+g^2_Y\over8}\left[
1-M^2_ZB^\prime_0(\nu,Z^0)+\xi_ZB_0(\nu,G^0)\right]
\label{eqn:Zcontr}
\end{eqnarray}
(we do not write any index on $\nu$ to stress that neutrino masses are set 
to zero in the $B_0$ functions). The diagram shown in Fig.~\ref{fig:ZGcontr}b, 
which arises due to the non-zero Majorana mass insertion, contributes to 
$\Sigma_m^{AB}(0)$. It gives
\begin{eqnarray}
I^{Z^0(2)}_{AB}=\delta^{AB}{g^2_2+g^2_Y\over2}\left[B_0(\nu,Z^0)
+{1\over4M_Z^2}a(G^0)-{1\over4M^2_Z}a(Z^0)\right]
\label{eqn:Zexot}
\end{eqnarray}
Finaly the exchange of $G^0$ and $\phi^0$ in the diagram shown 
in Fig.~\ref{fig:ZGcontr}c gives 
\begin{eqnarray}
I^{AB}_{G^0,\phi^0}=\delta^{AB}{1\over2v^2}\left[a(\phi^0)-a(G^0)\right].
\label{eqn:scalars}
\end{eqnarray}
where $16\pi^2a=m^2[-1+\ln(m^2/Q^2)]$ is another standard loop function.
The $\xi_Z$ dependent part of this contribution cancels
the $\xi_Z$ dependence of (\ref{eqn:Zexot}).

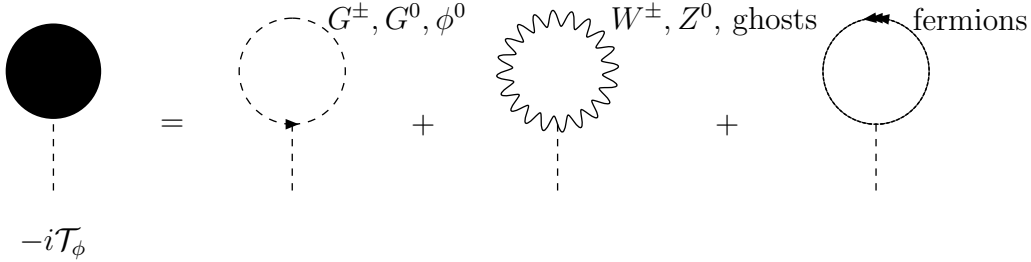
\begin{figure}[htbp]
\begin{center}
%\begin{tabular}{lp{280\unitlength}}
\begin{picture}(390,100)(0,0)
\DashLine(20,25)(20,50){3}
\Vertex(20,70){18}
\Text(20,5)[]{$-i{\cal T}_\phi$}
\Text(65,50)[]{$=$}
\DashLine(110,25)(110,50){3}
\DashArrowArc(110,70)(20,90,450){3}
\Text(150,90)[]{$G^\pm,G^0,\phi^0$}
\Text(160,50)[]{$+$}
\DashLine(210,25)(210,50){3}
\PhotonArc(210,70)(20,0,360){3}{20}
\Text(270,90)[]{$W^\pm,Z^0$, ghosts}
\Text(275,50)[]{$+$}
\DashLine(330,25)(330,50){3}
\DashArrowArc(330,70)(20,270,630){1}
\DashArrowArc(330,70)(20,277,637){1}
\DashArrowArc(330,70)(20,263,623){1}
\Text(367,90)[]{fermions}
\end{picture}
\end{center}
\caption{Tadpole diagrams.}
\label{fig:tadpoles}
%\end{tabular}
\end{figure}

The combined contribution of (\ref{eqn:GWcontr}-\ref{eqn:scalars}) is still
gauge dependent because one has to include the contribution of tadpole 
diagrams shown in fig.~\ref{fig:tadpoles}. They give
\begin{eqnarray}
&&I_{AB}^{\cal T}=-\delta^{AB}{{\cal T}_\phi\over M^2_\phi v}
= -\delta^{AB}{{\cal T}_\phi\over\lambda v^3}\nonumber\\
&&\phantom{aaaa}={\delta^{AB}\over16\pi^2}\left\{
{g^2\over4}\xi_W\left(1-\ln{\xi_WM^2_W\over Q^2}\right)+
{g^2+g^2_Y\over8}\xi_Z\left(1-\ln{\xi_ZM^2_W\over Q^2}\right)
\right.\nonumber\\
&&\phantom{aaaa}-{3\over2}\lambda\left(-1+\ln{M^2_\phi\over Q^2}\right)
+{1\over\lambda}\sum_{f,A}N^{(f)}_cy^4_{f_A}
\left(-1+\ln{m^2_{f_A}\over Q^2}\right)\label{eqn:tadpoles}\\
&&\phantom{aaaa}\left.
-{1\over\lambda}\left[{3\over8}g^4_2\left(-{1\over3}+\ln{M^2_W\over Q^2}\right)
+{3\over16}(g^2_2+g^2_Y)^2\left(-{1\over3}+\ln{M^2_Z\over Q^2}\right)
\right]\right\}\nonumber
\end{eqnarray}
It is easy to see that the $\xi_W$ and $\xi_Z$ dependence of 
(\ref{eqn:GWcontr}) and (\ref{eqn:Zcontr}) is canceled out by 
eq.~(\ref{eqn:tadpoles}).

To check that $\left(m_\nu^{\rm 1-loop}\right)^{AB}$ is
independent of the renormalization scale $Q$ we must recall the 
RGE for the vacuum expectation value $v^2$. Since in this approach
$v^2$ is merely an abbreviation for $-2m^2/\lambda$ where $m^2$ and $\lambda$
are the (negative) mass squared parameter and the self coupling of the Higgs
doublet, respectively, we have
\begin{eqnarray}
{d\over dt}v^2=v^2\left({1\over m^2}{d\over dt}m^2
-{1\over\lambda}{d\over dt}\lambda\right)
\label{eqn:rgevev}
\end{eqnarray}
where $16\pi^2t=\ln Q$. The RG equations for $m^2$ and $\lambda$ read:
\begin{eqnarray}
&&{d\over dt}m^2=m^2\left(-{9\over2}g^2_2-{3\over2}g^2_Y + 6\lambda +2T\right)
\nonumber\\
&&{d\over dt}\lambda=12\lambda^2-(9g^2_2+g^2_Y)\lambda
+ {9\over4}g^4_2 +{3\over2}g^2_2g^2_Y + {3\over4}g^4_Y
+4\lambda T-4T_2\phantom{aaa}
\nonumber
\end{eqnarray}
where $T\equiv\sum_fN^{(f)}_cy^2_f$ and
$T_2\equiv\sum_fN^{(f)}_cy^4_f$. Combining eq.~(\ref{eqn:rgevev}) with
the RGE (\ref{eqn:rge}) for $\mbox{\boldmath$C$}^{AB}$ we have therefore:
\begin{eqnarray}
&&{1\over4M}\mbox{\boldmath$C$}^{AB}(Q)v^2(Q) = 
{1\over4M}\mbox{\boldmath$C$}^{AB}(Q^\prime)v^2(Q^\prime)
-{3\over2}{v^2\over4M}\left(\mbox{\boldmath$C$}^{AB}y^2_{e_B}
+ y^2_{e_A}\mbox{\boldmath$C$}^{AB}\right)\ln{Q\over Q^\prime}
\nonumber\\  
&&\phantom{aaaaaaaaaaaaaaa}
-{v^2\over4M}\mbox{\boldmath$C$}^{AB}4\lambda\ln{Q\over Q^\prime}
+{v^2\over4M}\mbox{\boldmath$C$}^{AB}{3\over2}(g^2_2+g^2_Y)
\ln{Q\over Q^\prime} 
\nonumber\\  
&&\phantom{aaaaaaaaaaaaaaa}
- {1\over\lambda}\left({9\over4}g^4_2 +{3\over2}g^2_2g^2_Y + {3\over4}g^4_Y
-4T_2\right)\ln{Q\over Q^\prime} + \dots
\label{eqn:mtree_rge}
\end{eqnarray}

It is now easy to see that the leading $\ln Q$ dependence of the tree 
level neutrino mass matrix $\left(m_\nu^{\rm tree}\right)^{AB}(Q)$ 
cancels out with the explicit $\ln Q$ dependence of the 1-loop correction 
(\ref{eqn:Icordef}) to it.
In particular, the $1/\lambda$ terms in eq.~(\ref{eqn:mtree_rge})
cancel the $\ln Q$ dependence of the $W^\pm$, $Z^0$ (and their ghosts)
and fermionic tadpoles in (\ref{eqn:tadpoles}).

It is also possible to consider $v$ not as the tree-level VEV of the
Higgs field but instead as the minimum of the full 1-loop effective
potential. In such an approach there are no tadpoles
\footnote{Using $v$ determined from the full 1-loop potential is equivalent
to saying that one expands the symmetric Lagrangian around some initially 
unspecified $v$ and determines the value of $v$ from the requirement that 
the tree level tadpole (arising from a term in the Lagrangian that is linear 
in the Higgs field) cancels the 1-loop one.} 
but since the effective potential gives finite $v$ only in the Landau gauge, 
$\xi_W=\xi_Z=0$, the contributions (\ref{eqn:GWcontr}-\ref{eqn:scalars})
must be taken in this gauge too. In this approach the RGE for 
$v^2_{\rm 1-loop}$ is no longer given by eq.~(\ref{eqn:rgevev}) but instead 
is determined from the anomalous dimension (also taken in the Landau gauge)
of the Higgs field operator:
\begin{eqnarray}
{d\over dt}v^2_{\rm 1-loop} = 
v^2_{\rm 1-loop}\left({9\over4}g^2_2+{3\over2}g^2_Y-2T\right).
\end{eqnarray}
It is then easy to check that again the explicit $\ln Q$ dependence of
$\left(m_\nu^{\rm 1-loop}\right)^{AB}$ obtained from 
eqs.~(\ref{eqn:GWcontr}-\ref{eqn:scalars}) in the Landau gauge cancels
against the $\ln Q$ dependence of the tree level mass matrix
$(1/4M)\mbox{\boldmath$C$}^{AB}(Q)v^2_{\rm 1-loop}(Q)$. 

In practice, the difference between the two approaches (which formally
is a higher order effect) is not seen when $v^2$ (or $v^2_{\rm 1-loop}$)
is expressed in terms of the physical $Z^0$ boson mass. For example, in the 
first approach one has
\begin{eqnarray}
v^2={4(M^2_Z)_{\rm ph}\over g^2_2+g^2_Y}
\left[1-{\hat\Pi_{ZZ}(M^2_Z,Q)\over(M^2_Z)_{\rm ph}}
+{2{\cal T}_\phi\over\lambda v^3}\right]
\end{eqnarray}
where $\hat\Pi_{ZZ}(M^2_Z,Q)$ is the 1-PI self energy of the $Z^0$ boson
computed for $q^2=M^2_Z$ and 
renormalized in the $\overline{\rm MS}$ scheme with the renormalization 
scale $Q$. 

Neglecting terms of order ${\cal O}(m^4_{e_A}/M_W^4)$ and 
higher the final formula reads (tadpoles have canceled out) 
\begin{eqnarray}
&&I^{AB}={\delta^{AB}\over16\pi^2}\left\{
y^2_{e_B}\left({11\over8}-{3\over4}\ln{M^2_W\over Q^2}\right)
\right.\nonumber\\
&&\phantom{aaa}
+{g^2\over4}\left({1\over2}+\ln{M^2_W\over Q^2}\right)
+{g^2+g^2_Y\over8}\left(-{5\over2}+4\ln{M^2_Z\over Q^2}\right)
\label{eqn:sm_final}\\
&&\phantom{aaa}\left.
+{\lambda\over2}\left(-1+\ln{M^2_\phi\over Q^2}\right)-8\pi^2
{\hat\Pi_{ZZ}(M^2_Z,Q)\over(M^2_Z)_{\rm ph}}\right\}\nonumber
\end{eqnarray}
where we have adopted $\xi_{W,Z}=1$ and $v^2$ in 
the tree-level neutrino mass matrix is now given by 
\begin{eqnarray}
v^2\equiv {\hat s^2_W\hat c^2_W\over\pi\hat\alpha_{\rm EM}}(M^2_Z)_{\rm ph}
\nonumber
\end{eqnarray}
where $\hat s^2$ and $\hat\alpha_{\rm EM}$ are the sinus of the Weinberg 
angle and fine structure constant, respectively, in the $\overline{\rm MS}$
scheme and at the renormalization scale $Q$ (for which one can take 
$M_Z$).\footnote{In the standard way (see eg. \cite{POKBOOK}) $\hat s^2(M_Z)$ 
and $\hat\alpha_{\rm EM}(M_Z)$ can be expressed in terms of measurable 
quantities: $\alpha_{\rm EM}$ measured in the Thomson scattering and by the 
Fermi constant $G_F$.}
Note also that the factor $-{3\over2}\ln{M_W\over Q}$ in the first line of
(\ref{eqn:sm_final}) confirms the correctness of the recent re-derivation 
\cite{ANDRKELIRA} of the SM RGE.

{}From our discussion it should be clear that putting a particular emphasis 
on better stability with the renormalization scale $Q$ of the product
$v^2\mbox{\boldmath$C$}^{AB}$ (or of its eigenvalues) as in ref. \cite{SI}
makes no sense in the quantum field theory. The physical neutrino masses
defined as the poles of the propagators (or, in the one-loop approximation,
as the appropriate coefficients of the effective Lagrangian (\ref{eqn:leff}))
do not depend on the renormalization scale $Q$.
%: with one-loop accuracy, the 
%change of the values of the renormalized parameters of the SM (or MSSM) 
%Lagrangian with the scale $Q$ is, in measurable quantities, always 
%compensated for by explicit $Q$ dependence of one-loop corrections. 
On the other hand, by themselves big changes of $\mbox{\boldmath$C$}^{AB}$ 
during the RG evolution between $M_F$ and $M_Z$ do not signal any instability. 
They reflect only the importance of the resumation of large logarithmic 
contributions $\left(y^2_\tau\ln(M_F/M_Z)\right)^n$ where $n=1,2,\dots$  
%to all orders of the perturbative expansion 
in order to get reliable results
for neutrino masses in terms of the Lagrangian parameters defined at the
scale $M_F$. Finally, let us notice that $v^2(Q)$ disappears from the final
formula (\ref{eqn:sm_final}) for neutrino masses. Therefore, the question 
whether $v^2$ is considered as an abbreviation for $-m^2/\lambda$ (in which
case its variation with $Q$ is very rapid) or as the minimum of the full
one-loop effective potential is inessential for the stabilization of the
results for physical neutrino masses.

\section{The MSSM}
\setcounter{equation}{0}

In this section we calculate one-loop corrections to neutrino mass matrix in 
the MSSM. We will see that, apart from stabilizing
the results obtained from the RG analysis with respect to small changes of 
the final scale $Q$, they contain also $\ln Q$ independent terms which can be 
more important than the RG evolution. In some situations 
\cite{CHUPO,CHU,CHIOPOVA} they can change the pattern of mixing and lead to 
relations between the mixing angles different from the one obtained at the 
infrared fixed point of the RGE (\ref{eqn:rge}) \cite{CHKRPO}. We will use the
notation and conventions of ref.~\cite{ROS} in which the Feynman rules 
resulting from the renormalizable part of the MSSM (i.e. without the higher 
dimension operators) are collected. 

The neutrino masses arise in
supersymmetric models from one of the dimension 5 operators obtained by 
adding to the Lagrangian supersymmetric non-renormalizable terms
\begin{eqnarray}
&&\Delta{\cal L} \propto\mbox{\boldmath$C$}^{AB}\int d^2\theta
\left(\epsilon_{ij}\hat H^{(u)}_i\hat L^A_j\right)
\left(\epsilon_{lk}\hat H^{(u)}_l\hat L^B_k\right)+{\rm H.c.}
\phantom{aaa}\nonumber\\
&&\phantom{aaaa}=-{1\over4M}\mbox{\boldmath$C$}^{AB}\left[
\left(\epsilon_{ij} H^{(u)}_i l^A_j\right)
 \left(\epsilon_{ij} H^{(u)}_i l^B_j\right)
+\left(\epsilon_{ij} h^{(u)}_i L^A_j\right)
 \left(\epsilon_{ij} h^{(u)}_i L^B_j\right)\right.\label{eqn:dL_MSSM}\\
&&\phantom{aaaaaaaaaa}
\left.+2\left(\epsilon_{ij} H^{(u)}_i L^A_j\right)
 \left(\epsilon_{ij} h^{(u)}_i l^B_j\right)
+2\left(\epsilon_{ij} h^{(u)}_i L^A_j\right)
 \left(\epsilon_{ij} H^{(u)}_i l^B_j\right)\right] + {\rm H.c.}\nonumber\\
&&\phantom{aaaaaaaaaa}
+ {\rm terms ~with ~auxiliary ~fields}
\end{eqnarray}
where capital letters with a hat denote superfields and capital letters and
lower case letters denote their scalar and fermionic components, respectively.
In the second line we have fixed the normalization so that the first term
coincides with the operator (\ref{eqn:dL_SM}) in the SM.

Expressing the initial fields in terms of the physical ones as in 
ref.~\cite{ROS} one gets the following terms (we write down only those 
which are relevant for our calculation):
\begin{eqnarray}
&&\Delta {\cal L}_{\rm MSSM}=-{1\over2}(m_\nu^{\rm tree})^{AB}\nu_A\nu_B
\nonumber\\
&&\phantom{aaaaaaa}+{v_u\mbox{\boldmath$C$}^{AB}\over2\sqrt2M}
Z_H^{2k}H^+_ke_B\nu_A
-{v_u\mbox{\boldmath$C$}^{AB}\over4M}Z_R^{2k}H^0_k\nu_B\nu_A
\nonumber\\
&&\phantom{aaaaaaa}
-{\mbox{\boldmath$C$}^{AB}\over8M}Z_R^{2i}Z_R^{2j}H^0_iH^0_j\nu_B\nu_A
+{\mbox{\boldmath$C$}^{AB}\over8M}Z_H^{2i}Z_H^{2j}H^0_{i+2}H^0_{j+2}\nu_B\nu_A
\label{eqn:dL_MSSMint}\\
&&\phantom{aaaaaaa}
-{v_u\mbox{\boldmath$C$}^{AB}\over\sqrt2M}Z^{AJ}Z_N^{4i}\tilde\nu_J
\chi^0_i\nu_B
+{v_u\mbox{\boldmath$C$}^{AB}\over2\sqrt2M}Z_L^{Ak\ast}Z_+^{2j}
L^-_k\chi^+_j\nu_B+{\rm H.c.}\nonumber
\end{eqnarray}
As previously we use Weyl spinors here.
{}From eq.~(\ref{eqn:dL_MSSMint}) the necessary additional Feynman rules 
can be easily obtained.

The contribution to the quantity $I^{AB}$ defined in eq.~(\ref{eqn:Icordef})
of the $W^\pm$ and $G^\pm$ bosons is the same as in the SM and is given by
eq.~(\ref{eqn:GWcontr}).\footnote{Strictly speaking, in supersymmetry one 
has to use the DRED scheme \cite{CAJONI} instead of DIMREG used in the 
previous section. This would amount to omitting factors of 1 in the brackets 
in the second line of  eq.~(\ref{eqn:GWcontr}) and in eq.~(\ref{eqn:Zcontr})
and to similar changes in the tadpole contributions.
These changes do not affect, however, the interesting part of the 
threshold corrections which is not proportional to $\delta^{AB}$.}
The contribution of $H^\pm$ (arising from diagrams
similar to the ones shown in fig.~\ref{fig:cGcontr}) is
\begin{eqnarray}
&&I_{AB}^{H^\pm}=\delta^{AB}{g^2_2\over8}{m^2_{e_B}\over M^2_W}\tan^2\beta
\left[(m^2_{e_B}-M^2_{H^\pm})B^\prime_0(e_B,H^\pm)+B_0(e_B,H^\pm)\right]
\nonumber\\
&&\phantom{aaaaa}+\delta^{AB}{g^2_2\over2}{m^2_{e_B}\over M^2_W}
B_0(e_B,H^\pm).\phantom{aa}\label{eqn:cHcontr}
\end{eqnarray}
where $\tan\beta\equiv v_u/v_d$ is the usual ratio of the VEVs of the two
Higgs doublet $H^{(u)}$ and $H^{(d)}$.

The contribution of $Z^0$ is the same as in the SM and is given by 
eqs.~(\ref{eqn:Zcontr},\ref{eqn:Zexot}) while the contribution of neutral
scalars (\ref{eqn:scalars}) is in the MSSM replaced by
\begin{eqnarray}
I_{AB}^{\rm scalars} = \delta^{AB}{1\over2v^2_u}\left[\sin^2\alpha ~a(H^0) 
+ \cos^2\alpha ~a(h^0) - \cos^2\beta ~a(A^0)- \sin^2\beta ~a(G^0)\right]
\label{eqn:higgses}
\end{eqnarray}
Since $v^2_u/\sin^2\beta=v^2_u+v^2_d$ the last term of eq.~(\ref{eqn:higgses})
cancels the $\xi_Z$ dependence in eq.~(\ref{eqn:Zexot}) as in the SM.
The dependence of (\ref{eqn:GWcontr}) and (\ref{eqn:Zcontr}) on $\xi_W$ and
$\xi_Z$, respectively is again canceled out by the tadpole diagrams with
$G^\pm$ and $G^0$ loops.\footnote{In the MSSM eq.~(\ref{eqn:tadpoles}) is 
replaced by $-\delta^{AB}\left[{\cal T}_{h^0}\cos\alpha/M^2_h
+{\cal T}_{H^0}\sin\alpha/M^2_H\right]/v_u$. This can be brought to a more
convenient form by using the tree-level relations:
$\cos^2\alpha/M^2_h+\sin^2\alpha/M^2_H=
(\sin^2\beta/M^2_Z+\cos^2\beta/M^2_A)/\cos^22\beta$ and
$\sin\alpha\cos\alpha(1/M^2_h-1/M^2_H) = -(\sin\beta\cos\beta/\cos^22\beta)
(1/M^2_Z+1/M^2_A)$.}

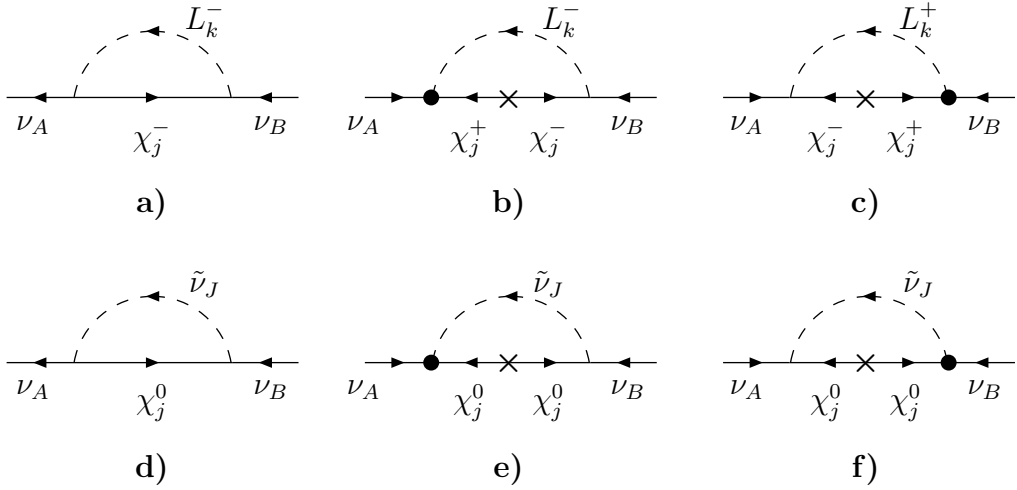
\begin{figure}[htbp]
\begin{center}
%\begin{tabular}{lp{280\unitlength}}
\begin{picture}(390,200)(0,0)
\ArrowLine(30,140)(5,140)
\ArrowLine(30,140)(90,140)
\ArrowLine(115,140)(90,140)
\DashArrowArc(60,135)(30,10,170){5}
\Text(15,130)[]{$\nu_A$}
\Text(105,130)[]{$\nu_B$}
\Text(60,125)[]{$\chi^-_j$}
\Text(80,170)[]{$L^-_k$}
\Text(60,100)[]{\bf a)}
\ArrowLine(140,140)(165,140)
\ArrowLine(195,140)(165,140)
\ArrowLine(195,140)(225,140)
\ArrowLine(250,140)(225,140)
\DashArrowArc(195,135)(30,10,170){5}
\Vertex(165,140){3}
\Text(195,140)[]{$\mbox{\boldmath$\times$}$}
\Text(140,130)[]{$\nu_A$}
\Text(240,130)[]{$\nu_B$}
\Text(180,125)[]{$\chi^+_j$}
\Text(210,125)[]{$\chi^-_j$}
\Text(215,170)[]{$L^-_k$}
\Text(195,100)[]{\bf b)}
\ArrowLine(275,140)(300,140)
\ArrowLine(330,140)(300,140)
\ArrowLine(330,140)(360,140)
\ArrowLine(385,140)(360,140)
\DashArrowArc(330,135)(30,10,170){5}
\Text(330,140)[]{$\mbox{\boldmath$\times$}$}
\Vertex(360,140){3}
\Text(285,130)[]{$\nu_A$}
\Text(375,130)[]{$\nu_B$}
\Text(315,125)[]{$\chi^-_j$}
\Text(345,125)[]{$\chi^+_j$}
\Text(350,170)[]{$L_k^+$}
\Text(330,100)[]{\bf c)}
\ArrowLine(30,40)(5,40)
\ArrowLine(30,40)(90,40)
\ArrowLine(115,40)(90,40)
\DashArrowArc(60,35)(30,10,170){5}
\Text(15,30)[]{$\nu_A$}
\Text(105,30)[]{$\nu_B$}
\Text(60,25)[]{$\chi^0_j$}
\Text(80,70)[]{$\tilde\nu_J$}
\Text(60,0)[]{\bf d)}
\ArrowLine(140,40)(165,40)
\ArrowLine(195,40)(165,40)
\ArrowLine(195,40)(225,40)
\ArrowLine(250,40)(225,40)
\DashArrowArc(195,35)(30,10,170){5}
\Vertex(165,40){3}
\Text(195,40)[]{$\mbox{\boldmath$\times$}$}
\Text(140,30)[]{$\nu_A$}
\Text(240,30)[]{$\nu_B$}
\Text(180,25)[]{$\chi^0_j$}
\Text(210,25)[]{$\chi^0_j$}
\Text(210,70)[]{$\tilde\nu_J$}
\Text(195,0)[]{\bf e)}
\ArrowLine(275,40)(300,40)
\ArrowLine(330,40)(300,40)
\ArrowLine(330,40)(360,40)
\ArrowLine(385,40)(360,40)
\DashArrowArc(330,35)(30,10,170){5}
\Text(330,40)[]{$\mbox{\boldmath$\times$}$}
\Vertex(360,40){3}
\Text(285,30)[]{$\nu_A$}
\Text(375,30)[]{$\nu_B$}
\Text(315,25)[]{$\chi^0_j$}
\Text(345,25)[]{$\chi^0_j$}
\Text(350,70)[]{$\tilde\nu_J$}
\Text(330,0)[]{\bf f)}
\end{picture}
\end{center}
\caption{Contributions of charginos/charged sleptons ({\bf a}-{\bf c})
and neutralinos/sneutrinos ({\bf d}-{\bf f}).}
\label{fig:susycontr}
%\end{tabular}
\end{figure}

Feynman diagrams describing contributions of chargino/charged slepton
and neutralino/sneutrino sectors are shown in fig.~\ref{fig:susycontr}.
They give:
\begin{eqnarray}
&&I_{AB}^{\rm charg}={1\over4}
\left(g_2Z_L^{Ak\ast}Z_-^{1j\ast}+y_{e_A}Z_L^{3+Ak\ast}Z_-^{2j\ast}\right)
\left(g_2Z_L^{Bk}Z_-^{1j}+y_{e_B}Z_L^{3+Bk}Z_-^{2j}\right)\nonumber\\
&&\phantom{aaaaaaaa}\times
\left[(m^2_{C_j}-M^2_{L_k})B^\prime_0(C_j,E_k^\pm)+B_0(C_j,E_k^\pm)\right]
\label{eqn:Ccontr}\\
&&\phantom{aaaaa}-{\sqrt2\over v_u}Z_L^{Ak\ast}Z_+^{2j}
\left(g_2Z_L^{Bk}Z_-^{1j}+y_{e_B}Z_L^{3+Bk}Z_-^{2j}\right)m_{C_j}
B_0(C_j,E_k^\pm)\nonumber
\end{eqnarray}
and 
\begin{eqnarray}
&&I_{AB}^{\rm neutr}={g^2_2+g^2_Y\over8}Z_\nu^{AJ}Z_\nu^{BJ\ast}
\left|s_WZ_N^{1j}-c_WZ_N^{2j}\right|^2
\left[(m^2_{N_j}-M^2_{\tilde\nu_J})B^\prime_0(N_j,\tilde\nu_J)+
B_0(N_j,\tilde\nu_J)\right]
\phantom{aaa}\nonumber\\
&&\phantom{aaaaa}-{2\over v_u}\sqrt{g^2_2+g^2_Y}Z_\nu^{AJ}Z_\nu^{BJ\ast}
Z_N^{4j}\left(s_WZ_N^{1j}-c_WZ_N^{2j}\right)m_{N_j}B_0(N_j,\tilde\nu_J)
\label{eqn:Ncontr}
\end{eqnarray}

To check that the $\ln Q$ dependence of the correction to the neutrino mass 
matrix in the MSSM matches the one following from the RGE it is more
convenient to use the second approach described in the previous section
and to assume that $v^2_u$ in the tree-level neutrino mass matrix is 
determined from the full 1-loop effective potential. Using then its RGE
\cite{GARIZW}
\begin{eqnarray}
{d\over dt}(v^2_u)_{\rm 1-loop}=(v^2_u)_{\rm 1-loop}
\left[{3\over2}g^2_2+ {1\over2}g^2_Y - 6\sum_Ay^2_{u_A}\right].
\end{eqnarray}
and the MSSM RGE for $\mbox{\boldmath$C$}^{AB}$ (\ref{eqn:rge}) one finds
\begin{eqnarray}
&&\left[{v^2_{\rm 1-loop}\over4M}\mbox{\boldmath$C$}^{AB}\right](Q) = 
\left[{v^2_{\rm 1-loop}\over4M}\mbox{\boldmath$C$}^{AB}\right](Q^\prime)
-{v^2_{\rm 1-loop}\over4M}\mbox{\boldmath$C$}^{AB}
\left({9\over2}g^2_2+{3\over2}g^2_Y\right)\ln{Q\over Q^\prime} 
\nonumber\\  
&&\phantom{aaaaaaaaaaaaaaa}
+{v^2_{\rm 1-loop}\over4M}\left(\mbox{\boldmath$C$}^{AB}y^2_{e_B}
+ y^2_{e_A}\mbox{\boldmath$C$}^{AB}\right)\ln{Q\over Q^\prime}
\label{eqn:treerge}
\end{eqnarray}
It is then easy to see that in the gauge $\xi_W=\xi_Z=0$ (in which the 
effective potential and hence also $v_{\rm 1-loop}$ is defined) the 
dependence on $\ln Q$ in the sum of corrections 
(\ref{eqn:GWcontr})-(\ref{eqn:Zexot}) and 
(\ref{eqn:cHcontr})-(\ref{eqn:Ncontr}) cancels out with 
the $\ln Q$ dependence in eq.~(\ref{eqn:treerge}).

The final formula for the factor $I^{\rm th}_{AB}$ in the MSSM takes the form
\begin{eqnarray}
&&16\pi^2I^{\rm th}_{AB}=\delta^{AB}{g^2\over2}{m^2_{e_B}\over M_W^2}
\left\{{1\over4}(1+\tan^2\beta)
\left(-{1\over2}+\ln{M^2_{H^\pm}\over Q^2}\right)
+{1\over2}\left(1+{3\over2}\ln{M^2_{H^\pm}\over M^2_W}\right)\right\}
\phantom{aaa}\nonumber\\
&&\phantom{aaaaaa}+{1\over4}
\left(g_2Z_L^{Ak\ast}Z_-^{1j\ast}+y_{e_A}Z_L^{3+Ak\ast}Z_-^{2j\ast}\right)
\left(g_2Z_L^{Bk}Z_-^{1j}+y_{e_B}Z_L^{3+Bk}Z_-^{2j}\right)\nonumber\\
&&\phantom{aaaaaaaaa}\times\left[\ln{M^2_{E^\pm_k}\over Q^2}+
f(m^2_{C_j},M^2_{E_k^\pm})\right]\nonumber\\
&&\phantom{aaaaa}-{\sqrt2\over v_u}Z_L^{Ak\ast}Z_+^{2j}
\left(g_2Z_L^{Bk}Z_-^{1j}+y_{e_B}Z_L^{3+Bk}Z_-^{2j}\right)m_{C_j}
\left[\ln{M^2_{E^\pm_k}\over Q^2}+g(m^2_{C_j},M^2_{E_k^\pm})\right]\nonumber\\
&&\phantom{aaaaa}+
{g^2_2+g^2_Y\over8}Z_\nu^{AJ}Z_\nu^{BJ\ast}
\left|s_WZ_N^{1j}-c_WZ_N^{2j}\right|^2
\left[\ln{M^2_{\tilde\nu_J}\over Q^2}+f(m^2_{N_j},M^2_{\tilde\nu_J})\right]
\phantom{aaa}\nonumber\\
&&\phantom{aaaaa}-{2\over v_u}\sqrt{g^2_2+g^2_Y}Z_\nu^{AJ}Z_\nu^{BJ\ast}
Z_N^{4j}\left(s_WZ_N^{1j}-c_WZ_N^{2j}\right)m_{N_j}
\left[\ln{M^2_{\tilde\nu_J}\over Q^2}+g(m^2_{N_j},M^2_{\tilde\nu_J})\right]
\nonumber\\
&&\phantom{aaaaaa}+{\rm terms ~proportional ~to} ~\delta^{AB}
\label{eqn:MSSMcor}
\end{eqnarray}
where the functions $f$ and $g$ are 
\begin{eqnarray}
&&f(a,b)= -{1\over2}+{a\over b-a}+{a^2\over(b-a)^2}\ln{a\over b}\phantom{aaa}
\nonumber\\
&&g(a,b)= -1-{a\over b-a}\ln{a\over b}\nonumber
\end{eqnarray}
and satisfy $f(a,a)=g(a,a)=0$. The terms proportional to the unit matrix are 
not interesting as they change only the overall scale of the neutrino masses
and do not influence the mixing angles.

Consider now the simplest limit $M_{H^\pm}= M_{L_k^\pm}= M_{\tilde\nu_J}
=m_{C_j}= m_{N_j}\equiv M_S$.
Using the solution (\ref{eqn:ellosol}) and writing for $Q\approx M_Z$ 
the factors $I^{\rm rg}_{e_A}$ as
\begin{eqnarray}
I_{e_A}=\exp\left(-\int_0^{t_Q}y^2_{e_A}(t^\prime)dt^\prime\right)
\approx\exp\left(-\int_0^{t_S}y^2_{e_A}(t^\prime)dt^\prime\right)
\times\left[1-{y^2_{e_A}\over16\pi^2}\ln{M_S\over Q}\right]
\end{eqnarray}
where $t_Q=(1/16\pi^2)\ln(M_F/Q)$, $t_S=(1/16\pi^2)\ln(M_F/M_S)$ 
and  $y^2_{e_A}=(g^2_2/2)(m^2_{e_A}/M^2_W)(1+\tan^2\beta)$,
we have (up to the overall normalization)
\begin{eqnarray}
&&\left(m_\nu^{\rm tree}\right)^{AB}(Q)\propto
\left(m_\nu^{\rm tree}\right)^{AB}(M_S)\\
&&\phantom{aaaa}
-{1\over16\pi^2}\ln{M_S\over Q}\left[
y^2_{e_A}\left(m_\nu^{\rm tree}\right)^{AB}(M_S)+
\left(m_\nu^{\rm tree}\right)^{AB}(M_S)y^2_{e_B}\right]
+{\cal O}\left(y^4_{e_A}\ln^2{M_S\over Q}\right)\nonumber
\end{eqnarray}
Adding the 1-loop correction in the form
\begin{eqnarray}
&&16\pi^2I^{AB}=\delta^{AB}{g^2\over2}{m^2_{e_B}\over M_W^2}
\left\{{1\over4}(1+\tan^2\beta)
\left(-{1\over2}+\ln{M^2_S\over Q^2}\right)
+{1\over2}\left(1+{3\over2}\ln{M^2_S\over M^2_W}\right)
\right\}\nonumber\\
&&\phantom{aaaaa}+{1\over4}\delta^{AB}y^2_{e_B}\ln{M_S\over Q}
+ {\rm terms ~proportional ~to} ~\delta^{AB}
\end{eqnarray}
we recover, as far the logarithms are concerned, the same result that is 
obtained with the running using the MSSM RGE from the scale $M_F$ down to 
$M_{\rm SUSY}$, and then the SM RGE to run down
to the $M_W$ scale \cite{CAESIBNA}. 
There is however a nontrivial extra non-logarithmic piece
\begin{eqnarray}
\delta^{AB}{g^2\over2}{m^2_{e_B}\over M_W^2}
\left({3\over8}-{1\over8}\tan^2\beta\right)\nonumber
\end{eqnarray}
which is missed by the usual procedure. This piece is usually less 
important than the effects of the evolution from the scale $M_F$ down 
to $M_{\rm SUSY}$ but for large $\tan\beta$ it is more important than 
the running from $M_{\rm SUSY}$  to the $M_W$ scale with the SM RGE.

\section{Numerical analysis}
\setcounter{equation}{0} 

In the SM the most important correction which changes the matrix structure 
of the neutrino mass matrix can be incorporated by substituting
\begin{eqnarray}
\mbox{\boldmath$C$}^{AB}(Q)\rightarrow\mbox{\boldmath$C$}^{AB}(Q)
+I_A^{\rm th}\mbox{\boldmath$C$}^{AB}(Q)
+\mbox{\boldmath$C$}^{AB}(Q)I_B^{\rm th}\label{eqn:corrs}
\end{eqnarray}
where
\begin{eqnarray}
I_A^{\rm th}={1\over16\pi^2}{g^2_2\over2}{m_{e_A}^2\over M^2_W}
\left[{11\over8}-{3\over2}\ln{M_W\over Q} 
+{\cal O}\left(x_A\ln x_A\right)\right]
\label{eqn:SMcor}
\end{eqnarray}
where $x_A\equiv m^2_{e_A}/M^2_W$. Since $I_A^{\rm th}$ are proportional
to the Yukawa couplings $y^2_{e_A}$, this correction cannot change 
qualitatively the results obtained by integrating the RGE and 
can be most easily taken into account by stopping the RG evolution of the
Wilson coefficient $\mbox{\boldmath$C$}^{AB}$ at the scale $Q=M_We^{-11/12}$.
The remaining corrections affect only the overall scale of the neutrino
masses and therefore are not interesting in view of the unspecified magnitude
of the mass $M$ in eq.~(\ref{eqn:treemass}). 

In the MSSM the contribution of $W^\pm$ and $H^\pm$ to $I^{\rm th}_{AB}$ 
is also proportional to $\delta^{AB}m^2_{e_A}/M^2_W$ and cannot change 
qualitatively the results of the RG evolution.\footnote{The same result
can be always obtained from the running by slightly changing the value of 
$\tan\beta$.}
However the effects of 
the genuinely supersymmetric contribution 
$I^{\rm susy}_{AB}=I^{\rm charg}_{AB}+I^{\rm neutr}_{AB}$ to 
$I^{\rm th}_{AB}$ can be important because unlike the SM case,
it is not necessarily proportional to $\delta^{AB}m^2_{e_A}/M^2_W$.
It has been demonstrated \cite{CHUPO,CHIOPOVA,CHU},
that $I^{\rm susy}_{AB}$, if large, can lead to relation between the 
mixing angles  different than the one obtained at the infrared
fixed point of the RGE \cite{CHKRPO}. The numerical estimates made in 
refs.~\cite{CHUPO,CHIOPOVA,CHU} relied however on approximating
the corrections $I^{\rm th}_{AB}$ by the pure wino contribution to 
$\Sigma_V^{AB}(0)$. Here we analyze the dependence of $I^{\rm th}_{AB}$ on
the MSSM parameters using the full expression (\ref{eqn:MSSMcor}). For
simplicity we will assume that the mixing of the left- and right-handed
charged sleptons is negligible.

\begin{figure}[htbp]
\begin{center}
\epsfig{file=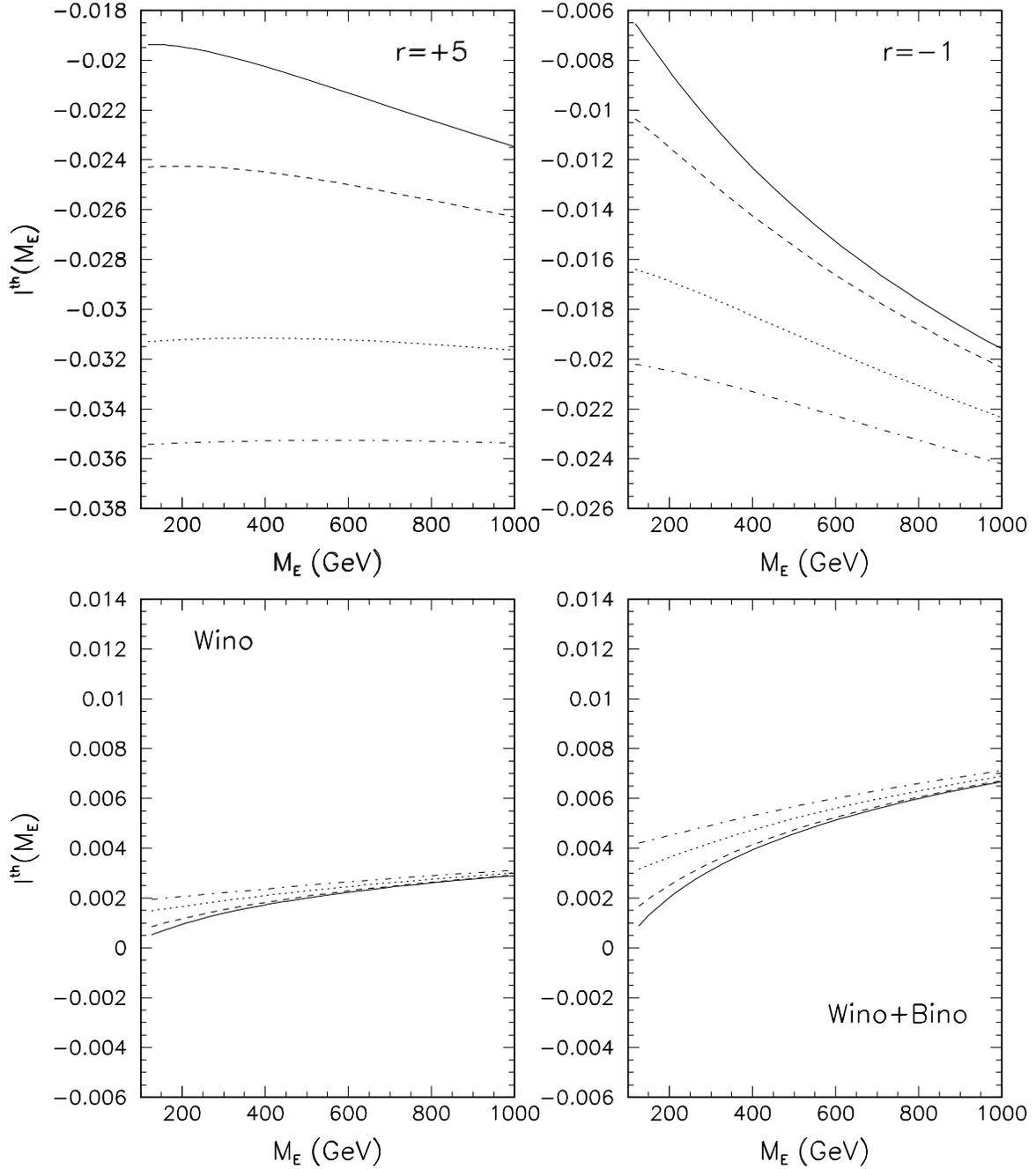,width=\linewidth}
\end{center}
\caption{Correction $I^{\rm susy}_L(M_E)$ as a function of the left-handed 
charged slepton mass for chargino mass 150 (solid), 250 (dashed), 500 
(dotted) and 800 (dot-dashed lines) for $\tan\beta=2$ and 
$r\equiv M_2/\mu=+5$ and $-1$. In lower panels we show the results of 
retaining only the wino (left) and wino and bino (right) contributions.}
\label{fig:thresh1}
\end{figure}

We begin by considering flavour conserving slepton mass matrices. In this 
case $I_{AB}^{\rm susy}=\delta^{AB}I_A^{\rm susy}$ because the matrices 
$Z_L^{Ak}$ and $Z_\nu^{AJ}$ are diagonal in the generation 
space. With no mixing of the left- and right-handed charged sleptons, the 
chargino and neutralino contribution can be simplified to
\begin{eqnarray}
&&16\pi^2I_A^{\rm susy}={1\over4}g_2^2\left|Z_-^{1j}\right|^2
\left[\ln{M^2_{E^\pm_{LA}}\over Q^2}
+f(m^2_{C_j},M^2_{E^\pm_{LA}})\right]\nonumber\\
&&\phantom{aaaaa}+
{g^2_2+g^2_Y\over8}\left|s_WZ_N^{1j}-c_WZ_N^{2j}\right|^2
\left[\ln{M^2_{\tilde\nu_A}\over Q^2}+f(m^2_{N_j},M^2_{\tilde\nu_A})\right]
\phantom{aaa}\nonumber\\
&&\phantom{aaaaa}-{\sqrt2\over v_u}g_2Z_+^{2j}
Z_-^{1j}m_{C_j}
\left[\ln{M^2_{E^\pm_{LA}}\over Q^2}
+g(m^2_{C_j},M^2_{E_{LA}^\pm})\right]\nonumber\\
&&\phantom{aaaaa}-{2\over v_u}\sqrt{g^2_2+g^2_Y}
Z_N^{4j}\left(s_WZ_N^{1j}-c_WZ_N^{2j}\right)m_{N_j}
\left[\ln{M^2_{\tilde\nu_A}\over Q^2}+g(m^2_{N_j},M^2_{\tilde\nu_A})\right]
\nonumber\\
&&\phantom{aaaaa}+
{1\over4}y^2_{e_A}\left|Z_-^{2j}\right|^2
\left[\ln{M^2_{E^\pm_{RA}}\over Q^2}+
f(m^2_{C_j},M^2_{E^\pm_{RA}})\right]
\label{eqn:corsimpl}
\end{eqnarray}
where $M_{E_{LA}^\pm}$ and $M_{E_{RA}^\pm}$ are the masses of the $A$-th
generation left- and right-handed charged sleptons, respectively. The 
contribution of the right-handed charged sleptons 
$\left(I_A^{\rm susy}\right)_R$ (last line of (\ref{eqn:corsimpl})) is again
proportional to the corresponding Yukawa coupling $y^2_{e_A}$. Hence, it too 
only slightly changes the effects of the RG evolution and can be neglected 
here. The remaining part $\left(I_A^{\rm susy}\right)_L$ of 
(\ref{eqn:corsimpl}) depends on the mass of the charged slepton (the mass of 
the sneutrino is related to it by the underlying $SU_L(2)$ symmetry: 
$M^2_{\tilde\nu_A}=M^2_{E^\pm_{LA}}+\cos2\beta M^2_W$), $\tan\beta$ and the 
parameters of the chargino/neutralino sector: $\mu$ and $M_2$ (as is customary,
in the neutralino sector we take $M_1\approx0.5M_2$). Figure~\ref{fig:thresh1}
shows the results of the numerical evaluation of 
$\left(I_A^{\rm susy}\right)_L\equiv I^{\rm susy}_L(M_{E^\pm_{LA}})$ and 
compares it to the results of approximating $I^{\rm susy}(M_{E^\pm_{LA}})$ by 
the pure $\tilde W^\pm$ (charged wino) or $\tilde W^\pm$, $\tilde W^0$ and 
$\tilde B$ (bino) 
contribution. The striking difference between the complete and approximate 
calculation is mainly due to the contribution of diagrams \ref{fig:susycontr}b,
c and e, f which give a negative contribution (third and fourth lines in 
eq.~(\ref{eqn:corsimpl})) but are missed in the approximation. Although the 
absolute magnitude of the the correction $I^{\rm susy}(M_{E^\pm_{LA}})$ does 
depend on $\tan\beta$ and the chargino composition, the differences 
$I^{\rm susy}(M_{E^\pm_{LA}})-I^{\rm susy}(M_{E^\pm_{LB}})$ (which are 
relevant 
for the changes of the neutrino mass matrix structure) are much less dependent
on $\tan\beta$. They are however sensitive to the chargino (and neutralino) 
composition as is clear from the comparison of the two upper panels 
of Fig.~\ref{fig:thresh1}. 

In ref.~\cite{CHIOPOVA} it  has been observed, that if\footnote{The 
renormalization group corrections (\ref{eqn:ellosol}), (\ref{eqn:lolaIfact})
can be incorporated in the formula (\ref{eqn:corrs}) by substituting
$I_A^{\rm th}\rightarrow I_A^{\rm th}-I_A^{\rm rg}$. One then has
\begin{eqnarray}
\mbox{\boldmath$C$}^{AB}(Q)\approx\mbox{\boldmath$C$}^{AB}(M_F)
+\left[I_A^{\rm th}(Q)-I_A^{\rm rg}(M_F/Q)\right]\mbox{\boldmath$C$}^{AB}(Q)
+\mbox{\boldmath$C$}^{AB}(Q)\left[I_B^{\rm th}(Q)-I_B^{\rm rg}(M_F/Q)\right].
\end{eqnarray}
}
$|I_e^{\rm susy}|\gg|I_\mu^{\rm susy}|$ and  
$I_\mu^{\rm susy} - (I_\tau^{\rm susy}-I_\tau^{\rm rg})\neq0$,
the masses of three neutrinos equal at the scale $M_F$ can be split
in agreement with the experimental information, provided the solar mixing 
angle is (very close to) maximal. For this mechanism to work 
$|I_e^{\rm susy}-I_\mu^{\rm susy}|\sim10^{-3}$ is required. We can now 
improve the estimates made in ref.~\cite{CHIOPOVA} on the basis of the wino 
approximation. From Fig.~\ref{fig:thresh1} it is clear that for 
$M_2/\mu\approx-1$ and lighter chargino mass $\sim150$ GeV the mass splitting 
$M_{\tilde\mu_L}\approx1.2M_{\tilde e_L}$ 
($M_{\tilde e_L}\approx1.2M_{\tilde\mu_L}$) is sufficient to obtain 
$I_e^{\rm susy}-I_\mu^{\rm susy}\sim10^{-3}$ ($\sim-10^{-3}$).
For heavier charginos and/or $M_2/\mu$ positive obtaining 
$|I_e^{\rm susy}-I_\mu^{\rm susy}|\sim10^{-3}$ requires very large mass 
splitting (or is even impossible to achieve). 

\begin{figure}[htbp]
\begin{center}
\epsfig{file=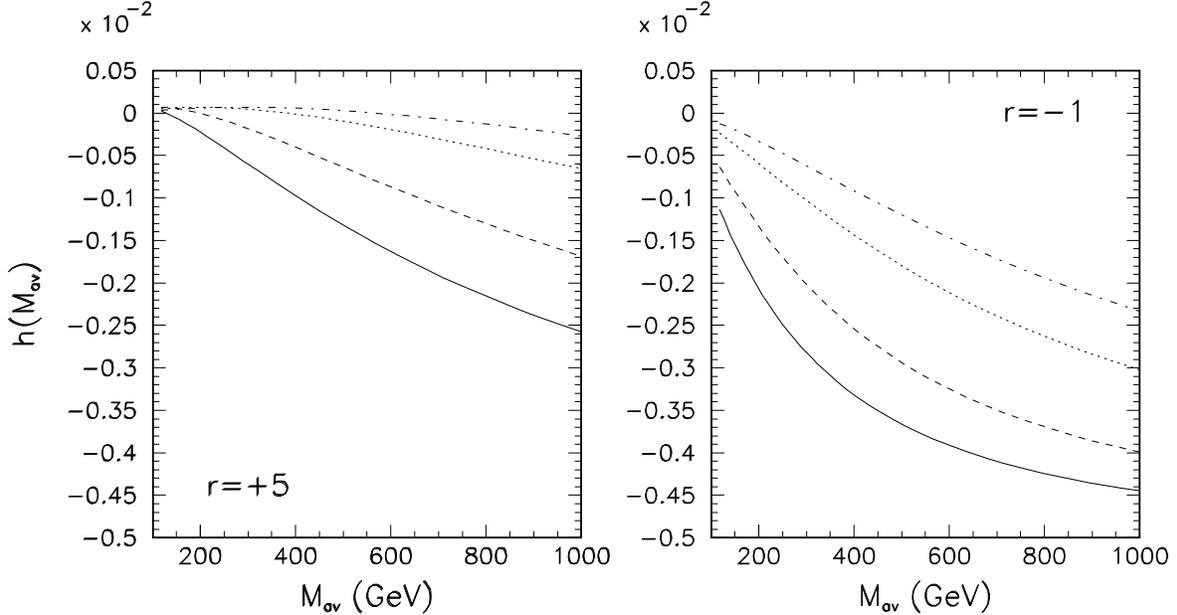,width=\linewidth}
\end{center}
\caption{Coefficient of the mass insertion $\delta^{AB}_{LL}$ 
as a function of the left-handed 
charged slepton mass for chargino mass 150 (solid), 250 (dashed), 500 
(dotted) and 800 (dot-dashed lines) for $\tan\beta=2$ and 
$r\equiv M_2/\mu=-1$ and $+5$.}
\label{fig:thresh2}
\end{figure}

{}From the formula (\ref{eqn:corsimpl}) we can also quantify the magnitude
of the off-diagonal corrections $I_{AB}^{\rm th}$ induced by the flavour 
mixing in the slepton mass matrices. Assuming 
%that the flavour violation is small and 
that left-handed charged slepton masses are all approximately
equal (which implies that sneutrino masses are also all approximately equal),
this is most easily done in the so-called mass insertion approximation 
\cite{GAGAMASI}. In eq.~(\ref{eqn:corsimpl}) terms involving e.g. charged 
left-handed sleptons can be written as
\begin{eqnarray}
&&Z_L^{Ak}Z_L^{Bk\ast}H(M^2_{E^\pm_k})=
Z_L^{Ak}Z_L^{Bk\ast}H(M^2_{\rm av}+(M^2_{E^\pm_k}-M^2_{E^\pm_{\rm av}}))
\nonumber\\
\phantom{aaaaaaa}&&
\approx\delta^{AB}\left[H(M^2_{E^\pm_{\rm av}})-M^2_{E^\pm_{\rm av}}
H^\prime(M^2_{E^\pm_{\rm av}})\right]
+Z_L^{Ak}M^2_{E^\pm_k}Z_L^{Bk\ast}H^\prime(M^2_{E^\pm_{\rm av}})
\nonumber\\
\phantom{aaaaaaa}&&
\approx\left(M^2_{LL}\right)_{AB}H^\prime(M^2_{E^\pm_{\rm av}})+
{\rm ~terms ~proportional ~to} ~\delta^{AB}
\end{eqnarray}
where $M_{\rm av}$ is the common mass of the left-handed charged sleptons and 
$H$ is some function. We have also used the defining property of
the matrices $Z_L$. The mass insertion are defined as the ratios of the 
off-diagonal elements $\left(M^2_{LL}\right)_{AB}$ of the mass squared matrix 
of sleptons to $M^2_{\rm av}$. In the case of flavour mixing only in the 
left-handed slepton sector\footnote{Flavour mixing in the right-handed 
slepton sector gives $I_{AB}^{\rm susy}$ suppressed by $y_{e_A}y_{e_B}$.
Flavour non-diagonal entries in the soft supersymmetry breaking terms 
mixing charged left- and right-handed sleptons give $I_{AB}^{\rm susy}$ 
proportional to $g_2y_{e_B}$ and, hence, substantial only for large 
$\tan\beta$ values, i.e. only when the renormalization group corrections are 
dominant.} 
we get for $I_{AB}^{\rm susy}$
\begin{eqnarray}
&&I_{AB}^{\rm susy}={\delta^{AB}_{LL}\over16\pi^2}
\left\{{1\over4}g_2^2\left|Z_-^{1j}\right|^2
F(m^2_{C_j},M^2_{E^\pm_{\rm av}})\right.\nonumber\\
&&\phantom{aaaaa}+
{g^2_2+g^2_Y\over8}\left|s_WZ_N^{1j}-c_WZ_N^{2j}\right|^2
F(m^2_{N_j},M^2_{\tilde\nu_{\rm av}})\nonumber\\
&&\phantom{aaaaa}-{\sqrt2\over v_u}g_2Z_+^{2j}
Z_-^{1j}m_{C_j}G(m^2_{C_j},M^2_{E^\pm_{\rm av}})\nonumber\\
&&\phantom{aaaaa}\left.-{2\over v_u}\sqrt{g^2_2+g^2_Y}
Z_N^{4j}\left(s_WZ_N^{1j}-c_WZ_N^{2j}\right)m_{N_j}
G(m^2_{N_j},M^2_{\tilde\nu_{\rm av}})\right\}
\end{eqnarray}
where
\begin{eqnarray}
&&F(a,b) ={b^2-3ab\over(b-a)^2}-{2a^2b \over(b-a)^3}\ln{a\over b}\nonumber\\
&&G(a,b)={b\over b-a} + {ab\over(b-a)^2}\ln{a\over b}.
\end{eqnarray}
We have used the fact that because of the underlying $SU_L(2)$ symmetry,
mass insertions in the left-handed charged slepton sector and in the
sneutrino sector are the same (and $M^2_{E^\pm_{\rm av}}$ and 
$M^2_{\tilde\nu_{\rm av}}$ are related). In fig.~\ref{fig:thresh2} we plot
the coefficient of $\delta^{AB}_{LL}$ as a function of $M^2_{E^\pm_{\rm av}}$
for several values of the chargino masses for $\tan\beta=2$.
We see that for a fixed $\delta^{AB}_{LL}$, the biggest values of
$I_{AB}^{\rm susy}$ are obtained for $M_2/\mu\approx-1$ and for rather
large slepton to chargino mass ratio. In principle the mass 
insertion approximation should fail for $|\delta_{XY}^{AB}|\simlt0.1$. 
In practice it works as an order of magnitude estimate even for 
$|\delta_{XY}^{AB}|\simlt1$ (the error is then of order 25\%).
More accurate results can be always obtained from the general formula
(\ref{eqn:MSSMcor}).

\section{Conclusions}

We have computed the low energy threshold corrections to neutrino masses
and mixings in the SM and in the MSSM. We have explicitly demonstrated that 
they are gauge independent and stabilize the results with respect to the 
variation of the scale $Q$ to which the relevant RGE is integrated from the 
high energy scale of the see-saw mechanism, thus clarifying the points
raised in ref.~\cite{SI}. 

The general formulae for the corrections $I_{AB}^{\rm th}$ derived in this 
paper can be applied to various models predicting the neutrino masses and 
mixing. They can be used to quantify the slepton mass splitting and/or the 
amount of flavour violation in the slepton sector necessary to realize the 
specific mechanisms, investigated in papers \cite{CHUPO,CHIOPOVA,CHU}, 
allowing to obtain correct mass squared differences and mixing angles from 
initially equal neutrino masses.  
They can find particularly interesting application in concrete models
\cite{BADUMO} relating the see-saw mechanism generating neutrino masses to 
flavour non-conservation in the slepton sector. Finally, they will be
indispensable for future precision tests of any quantitative theory of 
neutrino masses.

\vskip1.0cm
\noindent {\bf Acknowledgments}
\vskip 0.3cm
\noindent 
P.H.Ch. thanks S. Pokorski for interesting discussions.
His work was supported partially by the EC Contract 
HPRN-CT-2000-00148 for years 2000-2004 and by the Polish State 
Committee for Scientific Research grant 5 P03B 119 20 for years 
2001-2002.


\vskip1.0cm

\begin{thebibliography}{99}

\bibitem{WE} S. Weinberg, {\sl Phys. Rev. Lett.} {\bf 43} (1979), 1566.

\bibitem{CHPL} P.H. Chankowski and Z. P\l uciennik, {\sl Phys. Lett.} 
               {\bf B316} (1993), 312;
               K.S. Babu, C.N. Leung and J. Pantaleone, {\sl Phys. Lett.} 
               {\bf B319} (1993), 191. 

\bibitem{ANDRKELIRA} S. Antusch, M. Drees, J. Kersten, M. Lindner and
                     M. Ratz, preprint TUM-HEP-424/01 (hep-ph/0108005).

\bibitem{ELLO} J. Ellis and S. Lola, {\sl Phys. Lett.} {\bf B458} (1999), 310.

\bibitem{ALL} M. Tanimoto, {\sl Phys. Lett.} {\bf B360} (1995), 41;
              J. Ellis, G.K. Leontaris, S. Lola and D.V. Nanopoulos,
              {\sl Eur. Phys. J.} {\bf C9} (1999) 389;
              J.A. Casas, J.R. Espinosa, A. Ibarra and I. Navarro,
              {\sl Nucl. Phys.} {\bf B556} (1999), 3, {\bf B569} 
              (2000), 82, {\sl JHEP} {\bf 9909}:015 (1999);
              S. Lola, proceedings of the 6$^{th}$ Hellenic School and 
              Workshop on Elementary Particle Physics, Corfu, Greece, 
              September 1998 (hep-ph/9903203);
              N. Haba and N. Okamura, {\sl Eur. Phys. J.} 
              {\bf C14} (2000), 347;
              N. Haba, Y, Matsui and N. Okamura,
              {\sl Prog. Theor. Phys.} {\bf 103} (2000), 807;
              K.R.S. Balaji, A. Dighe, R.N. Mohapatra and M.K. Parida, 
              {\sl Phys. Lett.} {\bf B481} (2000), 33.

\bibitem{CHKRPO} P.H. Chankowski, W. Kr\'olikowski and S. Pokorski,
                 {\sl Phys. Lett.} {\bf B473} (2000), 109;
                 J.A. Casas, J.R. Espinosa, A. Ibarra and I. Navarro,
                 {\sl Nucl. Phys.} {\bf B573} (2000), 652.

\bibitem{CAESIBNA} J.A. Casas, J.R. Espinosa, A. Ibarra and I. Navarro,
                   {\sl Nucl. Phys.} {\bf B569} (2000), 82.

\bibitem{CHUPO} E. J. Chun and S. Pokorski, {\sl Phys. Rev.} {\bf D62}:053001
                (2000).

\bibitem{CHIOPOVA} P.H. Chankowski, A. Ioannisian, S. Pokorski and J.W.F.
                   Valle, {\sl Phys. Rev. Lett.} {\bf 86} (2001), 3488.

\bibitem{CHU} E. J. Chun, {\sl Phys. Lett.} {\bf B505} (2001) 155.

\bibitem{SI} N.N. Singh, {\sl Eur. Phys. J.} {\bf C19} (2001), 137.

\bibitem{POKBOOK} S. Pokorski, {\sl Gauge Field Theories}, 2nd edition,
                  Cambridge University Press, Cambridge, 2000, Chapters 1 and 
                  12. 

\bibitem{CAJONI} D.M. Capper, D.R.T. Jones and P. van Nieuwenhuizen,  
                 {\sl Nucl. Phys.} {\bf B167} (1980), 479.

\bibitem{ROS} J. Rosiek, {\sl Phys. Rev.} {\bf D41} (1990), 3464,
              {\sl Erratum} hep-ph/9511250.

\bibitem{GARIZW} G. Gamberini, G. Ridolfi and F. Zwirner, 
                 {\sl Nucl. Phys.} {\bf B331} (1990), 331.

\bibitem{GAGAMASI} F. Gabbiani, E. Gabrieli, A. Masiero and L. Silvestrini,
                   {\sl Nucl. Phys.} {\bf B447} (1996), 321; see also, M. 
                   Misiak, S. Pokorski and J. Rosiek, in {\sl Heavy Flavours 
                   II}, eds. A.J. Buras and M. Lindner, World Scientific
                   Publishing Co., Singapore 1998 ({\sl hep-ph}/9703442).

\bibitem{BADUMO} K.S. Babu, B. Dutta and R.N. Mohapatra, {\sl Phys. Lett.} 
                 {\bf B458} (1999), 93; W. Buchm\"uller, D. Delepine and F. 
                 Vissani, {\sl Phys. Lett.} {\bf B459} (1999), 171; W. 
                 Buchm\"uller, D. Delepine and L.T. Handoko, {\sl Nucl. Phys.}
                 {\bf B576} (2000), 445; J. Ellis, M.E. Gomez, G.K. Leontaris, 
                 S. Lola and D.V. Nanopoulos, {\sl Eur. Phys. J.} {\bf C12} 
                 (2000), 319; J. Hisano and K. Tobe, {\sl Phys. Lett.} 
                 {\bf B510} (2001), 197; D.F. Carvalho, J. Ellis and S. Lola, 
                 {\sl Phys. Lett.} {\bf B515} (2001), 323; T. Bla\v zek 
                 and S.F. King, {\sl Phys. Lett.} {\bf B518} (2001), 109;
                 S. Lavignac, I. Masina and C.A. Savoy, preprint SACLAY-T01/066
                 (hep-ph/0106245).

\end{thebibliography}
\end{document}